\documentclass[aps,prd,12pt,preprint,nofootinbib,superscriptaddress,tightenlines,floatfix,showpacs]{revtex4}
 
\usepackage{graphicx}
\usepackage{rotating}
\usepackage{dcolumn}
\usepackage{bm}
\usepackage{longtable}

\begin{document}
\newcommand{\ie}{{\it i.e.}}
\newcommand{\eg}{{\it e.g.}}
\newcommand{\etap}{{\eta\,'}}
\newcommand{\psip}{{\psi(2S)}}
\newcommand{\dipi}{{\pi^+\pi^-}}
\newcommand{\jpsi}{{J/\psi}}
\newcommand{\piz}{{\pi^0}}

\newcommand{\SB}{side band}
\newcommand{\XF}{cross feed}
\newcommand{\ipb}{pb$^{-1}$}
\newcommand{\DM}{\delta}
\newcommand{\aDM}{\langle\DM\rangle}
\newcommand{\aDMi}{\langle\DM_i\rangle}
\newcommand{\aDMm}{\langle\DM_\dimu\rangle}
\newcommand{\aDMp}{\langle\DM_\dipi\rangle}
\newcommand{\bias}{\beta}
\newcommand{\sccont}{continuum subtraction}
\newcommand{\stavg}{(\aDM-\bias)_{\rm s}}
\newcommand{\etal}{{\sl et al.}}

\newcommand{\gga}{\gamma\gamma}
\newcommand{\tpi}{\pi^+\pi^-\pi^0}
\newcommand{\tpz}{3\pi^0}
\newcommand{\ppg}{\pi^+\pi^-\gamma}
\newcommand{\eeg}{e^+e^-\gamma}
\newcommand{\meta}{M_{\eta}}
\newcommand{\metapdg}{\meta^{\rm *}}
\newcommand{\metamc}{\meta^{\rm MC}}
\newcommand{\metap}{M_{\etap}}
\newcommand{\metapz}{M_{\etap}^0}
\newcommand{\metappdg}{\metap^{\rm *}}
\newcommand{\metapmc}{\metap^{\rm MC}}
\newcommand{\inv}{invisible}

\newcommand{\Mpp}{M_\psip}
\newcommand{\Mjp}{M_\jpsi}

\newcommand{\Mpppdg}{M^{\rm *}_\psip}
\newcommand{\Mjppdg}{M^{\rm *}_\jpsi}
\newcommand{\Mpizpdg}{M^{\rm *}_\piz}

\preprint{CLNS~08/2030}       
\preprint{CLEO~08-13}         

\title{\Large \boldmath
Measurement of the $\etap$-meson mass using
$\jpsi\to\gamma\etap$}

\author{J.~Libby}
\author{L.~Martin}
\author{A.~Powell}
\author{G.~Wilkinson}
\affiliation{University of Oxford, Oxford OX1 3RH, UK}
\author{K.~M.~Ecklund}
\affiliation{State University of New York at Buffalo, Buffalo, New York 14260, USA}
\author{W.~Love}
\author{V.~Savinov}
\affiliation{University of Pittsburgh, Pittsburgh, Pennsylvania 15260, USA}
\author{H.~Mendez}
\affiliation{University of Puerto Rico, Mayaguez, Puerto Rico 00681}
\author{J.~Y.~Ge}
\author{D.~H.~Miller}
\author{I.~P.~J.~Shipsey}
\author{B.~Xin}
\affiliation{Purdue University, West Lafayette, Indiana 47907, USA}
% ...
\collaboration{CLEO Collaboration}
\noaffiliation

\date{June 13, 2008}

\begin{abstract}
We measure the mass of the $\etap$ meson 
using $\psip$$\to$$\dipi\jpsi$, $\jpsi\to\gamma\etap$ 
events acquired with the CLEO-c detector
operating at the CESR $e^+e^-$ collider.
Using three decay modes, 
$\etap$$\to$$\rho^0\gamma$,
$\etap$$\to$$\dipi\eta$ with $\eta$$\to$$\gga$, and 
$\etap$$\to$$\dipi\eta$ with $\eta$$\to$$\dipi\piz$,
we find $\metap$=957.793$\pm$0.054$\pm$0.036~MeV, in which
the uncertainties are statistical and systematic, respectively. 
This result is consistent with but substantially more
precise than the current world average.

\end{abstract}

\pacs{14.40.Cs, 14.40.Aq}

\maketitle

Experimental precision on the $\etap$ mass is currently
worse than that of $\pi$, $K$, $\eta$, 
$\omega$, or $\phi$~\cite{PDG2008}.
The PDG world-average value~\cite{PDG2008} is 
$\metap=957.66\pm0.24~{\rm MeV}$.
Recent experimental focus on the $\eta$ mass
has resolved a conflict among discrepant measurements~\cite{PDG2008};
the $\etap$ mass uncertainty
now stands out in comparison to 
other narrow light mesons.
This Letter presents a new
measurement of the $\etap$ mass,
the first in more than a decade.

  The $\eta$ and $\etap$ mesons are commonly understood 
as mixtures of the pure SU(3)-flavor octet ($u\bar{u}+d\bar{d}$)
and singlet ($s\bar{s}$) states
with possible admixtures of gluonium~\cite{rosner,gilman}.
The strengths of pseudoscalar and gluonium mixing 
become manifest in ratios of branching fractions
for radiative decays of pseudoscalar ($P$) and vector ($V$) 
mesons, $V\to\gamma P$ and $P\to\gamma V$~\cite{besmix,kloemix,escrib}.
However, these effects should also be evident in relationships
between {\sl masses} of $\eta$ and $\etap$ on the one hand, 
and those of $\pi$ and $K$ mesons on the other.
In the {\sl current} (as opposed to {\sl constituent})
quark framework, one such formulation 
for the pseudoscalar mixing angle
in the flavor basis, $\phi_P$, finds~\cite{jones},
to first order in flavor-symmetry breaking~\cite{feldmann},
\begin{equation}
\tan^2\phi_P = { {(\metap^2 - 2M_K^2 + M_\pi^2)(\meta^2 - M_\pi^2)}
               \over{ (2M_K^2 - M_\pi^2 - \meta^2)(\metap^2 - M_\pi^2)}
               }\ .
\end{equation}
\noindent 
Using PDG~\cite{PDG2008} values for masses,
$\phi_P$=(41.460$\pm$0.009)$^\circ$,
which has an 
uncertainty dominated by $\Delta M_K$ ($\pm$0.007$^\circ$),
followed by $\Delta \metap$ ($\pm$0.004$^\circ$) 
and $\Delta\meta$ ($\pm$0.003$^\circ$).
This value for $\phi_P$ is consistent with 
determinations based upon branching fractions 
(which have uncertainties at the level of 
1$^\circ$~\cite{besmix,kloemix,escrib}),
indicating flavor-symmetry breaking effects are small~\cite{feldmann}.
In general, precision on most other predictions involving $\metap$
has not yet matched that of experiment,
and instead appears to be limited by theoretical assumptions
and approximations~\cite{kekez,gerard,michael}. 
More precise $\etap$ mass measurements will
act as grounding for such predictions as they evolve.

This measurement is based upon data acquired with
the CLEO detector at the CESR (Cornell Electron 
Storage Ring) symmetric $e^+e^-$ collider, 
mostly in the CLEO-c configuration (95\%) with the balance
from CLEO~III. The data sample
corresponds to 27$\times$10$^6$~\cite{xnext} 
produced $\psip$ mesons, of which about  4$\times$10$^4$
decay as $\psip$$\to$$\dipi\jpsi$, $\jpsi\to\gamma\etap$. 
We employ three decay modes, denoted as
$A$ ($\etap$$\to$$\rho^0\gamma$),
$B$ ($\etap$$\to$$\dipi\eta$ with $\eta\to\gga$), and
$C$ ($\etap$$\to$$\dipi\eta$ with $\eta\to\dipi\piz$).
The decay $\jpsi\to\gamma\eta$ with $\eta\to\tpi$,
denoted as mode $D$, is used to cross-check the
analysis method. 
Other decay modes of $\etap$ and $\eta$ were studied
and found to have inadequate statistical precision
compared to modes $A$-$D$.
Because both $\psip$ and $\jpsi$ are very narrow
resonances with precisely known masses 
($\jpsi$, $\pm$11~keV; $\psip$, $\pm$34~keV~\cite{PDG2008}),
imposition of kinematic constraints enables
a significant improvement in $\etap$ mass
resolution. 
The analysis method and systematic error considerations
are very similar to those of Ref.~\cite{cleometa},
in which $\psip\to\eta\jpsi$, $\jpsi\to\ell^+\ell^-$
decays were used to measure $M_\eta$. 

The CLEO-c detector is described in detail elsewhere~\cite{CLEO};
it offers 93\% solid angle coverage of precision 
charged particle tracking and an electromagnetic calorimeter
comprised of CsI(Tl) crystals.
The tracking system enables momentum measurements
for particles with momentum
transverse to the beam exceeding 50~MeV/$c$
and achieves resolution $\sigma_p/p\simeq0.6\%$ at $p$=1~GeV/$c$.
The barrel calorimeter reliably measures photon showers down
to $E_\gamma$=30~MeV and has a resolution of
$\sigma_E/E\simeq$5\% at 100~MeV and 2.2\% at 1~GeV.

Signal and background processes are modeled 
with Monte Carlo (MC) samples that were generated using the
{\sc EvtGen} event generator~\cite{evtgen},
fed through a {\sc Geant}-based~\cite{geant} detector simulation,
and subjected to event selection criteria. 
The distribution of ``transition dipion'' (the $\dipi$
emitted in the $\psip$-to-$\jpsi$-transition) mass is
sculpted~\cite{xnext} to match that of the data,
and angular distributions of the $\jpsi$ decay products
are set to be appropriate for a vector decay into
a vector and a pseudoscalar.
The decay $\eta\to\dipi\piz$ is generated using
the matrix element measured in Ref.~\cite{layter}.

Event selection requires the tracking system to find exactly two 
oppositely-charged particles for the transition dipion,
and two ($A$, $B$, and $D$) or four ($C$) more tracks of net charge zero.
A minimum of two ($A$) or three ($B$, $C$, and $D$) 
photon candidates must also be found. 
Photons must
be located in the central portion of the barrel calorimeter
where the amount of material traversed is smallest
and therefore energy resolution is best ($|\cos\theta_\gamma|$$<$0.75,
where $\theta$ is the polar angle with respect to the initial
$e^+$ direction).
Such photons must also have energy exceeding 120~MeV ($A$) 
or 36~MeV ($B$, $C$, and $D$), and either be more
than 30~cm from any shower associated with one of the charged pions,
or, when between 15 and 30~cm from such a shower,
have a photon-like lateral shower profile. Selected photons cannot
lie near the projection of any charged pion's trajectory into
the calorimeter, or align with the initial momentum 
of any $\pi^\pm$ candidate within 100~mrad.
   Spurious showers faking photons can result from
   nuclear interactions of charged pions in the calorimeter.
   Such showers tend to have low
   energy for which the MC modeling may be less accurate.
   Therefore we consider only the two ($A$) or 
   three ($B$, $C$, and $D$) highest energy photon candidates satisfying
   the above criteria to suppress such mismodeling effects.
Photon pairs are candidates for a $\piz$ or $\eta$ if their
invariant mass satisfies $M(\gga)$=115-150~MeV or 500-580~MeV, respectively, 
and are then constrained
to the PDG average $\piz$ or $\eta$ masses~\cite{PDG2008}.

\begin{figure}[t]
\includegraphics*[width=6.5in]{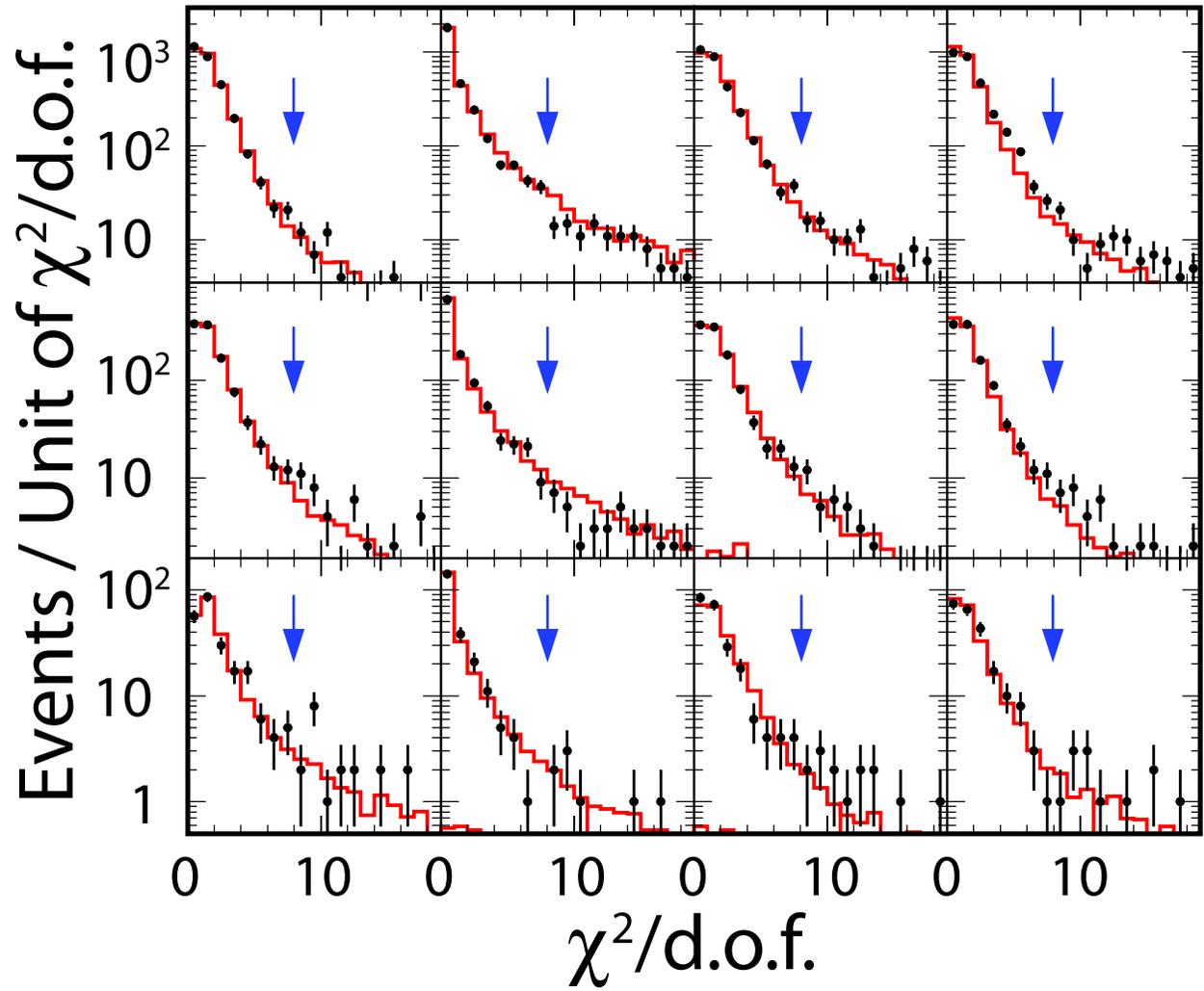}
\caption{Distributions of the fit quality for
the kinematic constraint procedures described in the text,
for, from top to bottom, modes $A$, $B$, and $C$,
and, from left to right, $\jpsi$ vertex fit, $\jpsi$ mass fit,
$\psip$ vertex fit, and $\psip$ four-momentum fit.
Arrows indicate nominal selection criteria.
The solid line histogram represents the signal MC prediction
normalized to the number of events in the data.
\label{fig:chisq} }
\end{figure}

  Further kinematic requirements are applied in two two-step fits: 
first, the $\jpsi$ decay products are constrained to originate 
from a single point (vertex) consistent with the beam spot 
and then constrained to have the $\jpsi$ mass, 
$\Mjp$~\cite{PDG2008}; quality restrictions are applied to 
both the vertex and the mass kinematic fits ($\chi^2/$d.o.f.$<$8 for each). 
Second, the $\jpsi$ candidate and transition dipion are
constrained to a common vertex and then to the $\psip$ mass, 
$\Mpp$~\cite{PDG2008}, and three-momentum, including the
effect of the $\simeq$3~mrad crossing angle of the $e^+$ and $e^-$ beams.
Again, quality restrictions are applied to both the
vertex and four-momentum kinematic fits ($\chi^2/$d.o.f.$<$8 for each). 
The $\chi^2/$d.o.f. distributions are shown in Fig.~\ref{fig:chisq}.

No attempt is made to isolate the $\eta\to\dipi\piz$
decay in mode $C$ because this is difficult to achieve in an unambiguous
way; frequently multiple such combinations per event are
consistent with an $\eta$ decay due to confusion between
the dipion from the $\etap-\eta$ transition and the dipion from
the $\eta$ decay. Therefore we do not take advantage
of constraining any three-pion mass to that of the $\eta$
as we do with $\eta\to\gga$ in mode $B$. The $\etap$ mass is
instead formed from the five pion four-momenta.

\begin{figure}[t]
\includegraphics*[width=6.5in]{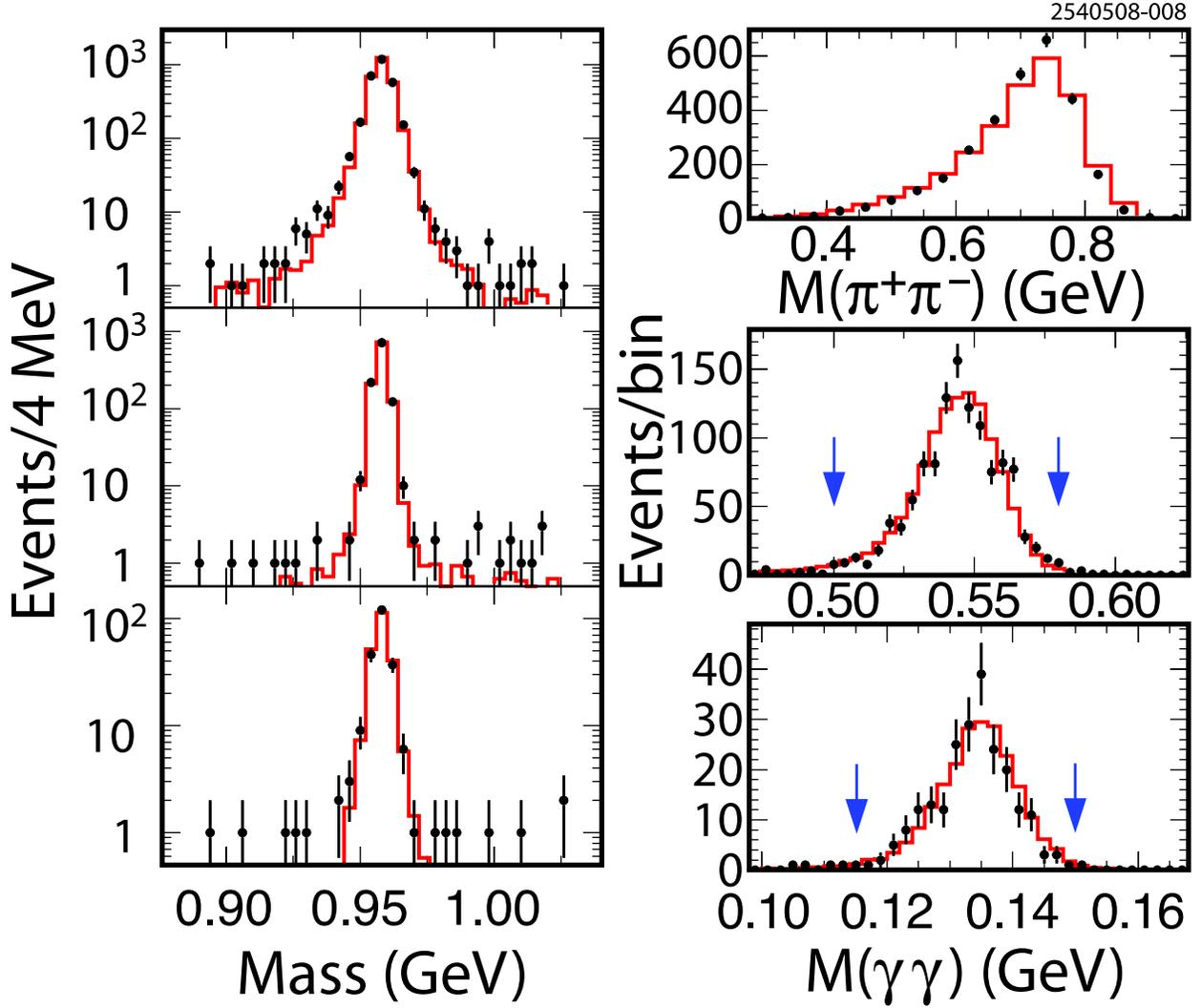}
\caption{Distributions of $\etap$-candidate
masses (left) for modes A (top), B (middle),
and C (bottom). On the right, masses of intermediate
particles: the $\rho^0\to\dipi$ mass for mode $A$, 
the $\eta\to\gga$ mass for mode $B$, and the $\piz\to\gga$
mass for mode $C$. Symbols are defined as in Fig.~\ref{fig:chisq}.
\label{fig:masses} }
\end{figure}

With these selections, the three samples
of $\etap$ decays are very pure.
Non-$\jpsi$ decays are estimated to constitute up to a 0.5-1.0\%
background for all three $\etap$ modes, as determined by
examining the recoil mass sidebands of the $\psi(2S)$ transition dipion 
for non-$\jpsi$
contamination when the $\jpsi$ vertex and mass fits are removed.
Inspection of the $\rho^0$, $\piz$, $\eta$, and $\etap$ mass distributions,
shown in Fig.~\ref{fig:masses}, verify that $\jpsi$ backgrounds are small.
MC studies of $\jpsi$ decays that might be expected
to contaminate mode $A$, such as $\jpsi\to\rho^0\eta$ and
$\jpsi\to\dipi\piz$ and cross-feed from $\jpsi\to\gamma\etap$,
$\etap\to\dipi\eta$, $\eta\to\gga$, are found to contribute 
only at the 0.1\% level. These backgrounds show no significant structure or
strong slope in the $\etap$ mass distributions,
so their effects upon the mass determination are 
small compared with statistical uncertainties
and are therefore neglected.

\begin{figure}[thp]
\includegraphics*[width=5.0in]{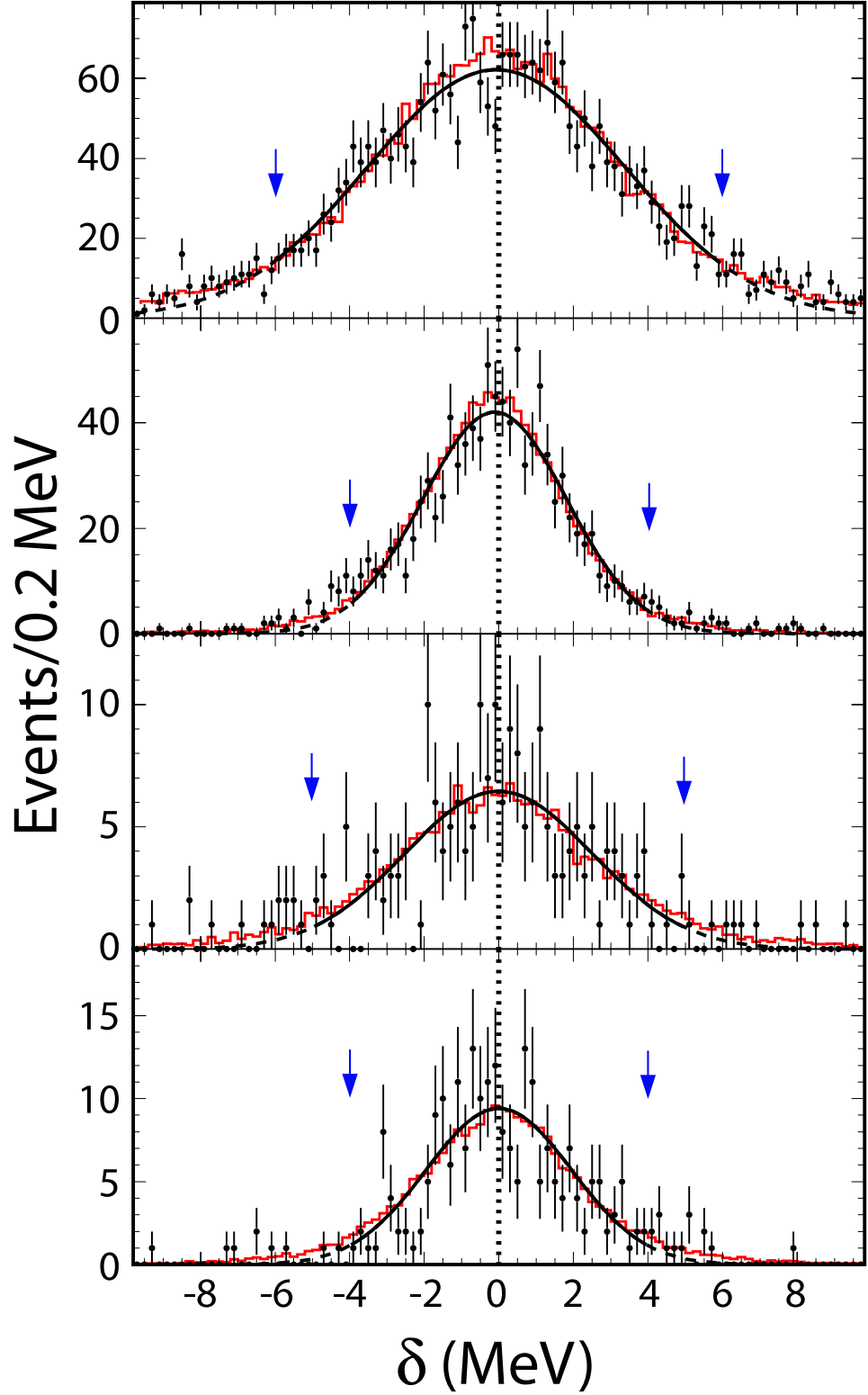}
\caption{Distributions of $\DM$ for, from top to bottom, modes
$A$, $B$, $C$, and $D$ (see text),
with the data represented by the points with error bars
and the Gaussian fit overlaid. 
The solid line
portion of the Gaussian curve indicates the 
mass window used for the fit
and the dashed portions its extension.
The solid line histogram represents
MC simulation of signal,
normalized as in Figs.~1 and 2. 
\label{fig:fit} }
\end{figure}

  Each event yields an invariant mass $M$ of the 
kinematically-constrained decay products;
a single mass value is extracted for each decay mode $i$
by fitting a Gaussian shape to the distribution of
$\DM_i$$\equiv$$M_i$$-$$M_0$,
where $M_0$ is a reference value, either the
PDG2006 world-average $\metapz=957.78~{\rm MeV}$~\cite{PDG2006}
(for historical reasons; the PDG2008 value is
$957.66\pm 0.24$~MeV~\cite{PDG2008}),
or, in the case of mode $D$, 
the PDG2008 $\eta$ mass central value. We define the symbols
$\aDM$ and $\sigma$ as the resulting Gaussian mean 
and width, respectively. The fits are restricted to
the central portion of each $\DM$ distribution
because the tails outside this region are not represented well by
a single Gaussian form. The fits
span $\pm$1.7$\sigma$ to $\pm$2.1$\sigma$
about $\aDM$, and in all cases
the resulting fit has a confidence level exceeding 40\%.
The distributions of $\DM_i$ 
with overlaid fits are shown
in Fig.~\ref{fig:fit}.
Other shapes that might fit the tails,
such as a double Gaussian, have been found to
yield unstable fits and/or do not improve precision
of finding the peak.

   There is an unavoidable 
low-side tail in any monochromatic photon
energy distribution from the CLEO calorimeter. 
It originates from losses sustained 
in interactions prior to impinging
upon the calorimeter and from leakage
outside those crystals used
in the shower reconstruction.
This asymmetric photon energy resolution
function causes 
a small but significant systematic bias
in $\aDM$: for simplicity
of the kinematic fitting formalism,
input uncertainties  are
assumed to be symmetric,
and a bias occurs if they are not. 
This bias in fitted Gaussian mean
is mode-dependent because each
presents a different mix of charged and neutral
particles.

The biases $\bias_i$ are estimated by following
the above-described procedure on MC signal samples.
Each $\bias_i$ is the difference between
the Gaussian peak value of the $\metap$ distribution
and the input $\metapmc$. We define the bias as
$\bias_i$$\equiv$$\aDMi_{\rm MC}$,
in which we use
the MC input $\metapmc$ for $M_0$.
A non-zero value of $\bias_i$ means that, for mode $i$,
the Gaussian peak mass $\aDMi$ is offset
from the true mass and must be corrected.
Although the asymmetric photon lineshape is
responsible for most of this difference,
the resulting correction automatically compensates for all
modeled sources of bias.
A bias value for the
$\eta$ cross-check is determined similarly.

  Table~\ref{tab:tableres} summarizes results by decay mode.
The $\eta$ mass result from mode $D$ ($\aDM - \bias =38\pm148$~keV) 
is consistent with expectations 
within its statistical uncertainty.
The number of events
involved in the determination of $\metap$ in modes
$A$, $B$, and $C$ is 3917.
The three values of $\aDMi-\bias_i$ have an
average, weighted by statistical errors only,
of $\stavg$=15$\pm$53~keV with a $\chi^2$=0.14
for two degrees of freedom, demonstrating 
consistency.

  Systematic and statistical errors are summarized in Table~\ref{tab:tablesys}.
Uncertainties that are uncorrelated mode-to-mode,
including statistical, are used to determine the weights 
($w_i$=0.44, 0.47, and 0.09 for $A$, $B$, and $C$, respectively) 
applied to combine values from the three modes into the weighted sum 
$(\aDM$$-$$\bias)_{\rm
  w}$=$\sum_{i=1}^{3}$$w_i$$\times$$(\aDMi$$-$$\bias_i)$=13$\pm$54$\pm$36~keV,
where the uncertainties are statistical and systematic, respectively.
Note that the combined value using weights including
systematic errors is virtualy identical to
that obtained accounting only for statistical uncertainties.

  As the mass distributions are not
perfectly Gaussian, there is some systematic
variation of the peak value with the choice
of mass window for each fit. We vary
the low- and high-side mass limits by $\pm$1~MeV,
symmetrically, and note the variations
in MC peak values with respect to the nominal mass window, 
as summarized in Table~\ref{tab:tablesys}.

Uncertainties attributable
to imprecision in the masses of the $\jpsi$
(11~keV), $\psip$ (34~keV), and $\eta$ (24~keV) mesons~\cite{PDG2008}
are directly calculated by repeating the analysis 
using an altered mass 
and the deviation in $\aDM$ per ``1$\sigma$'' change from nominal 
taken as the error. 
Based on the studies in Ref.~\cite{cleometa} 
we take one third of the bias magnitude plus
its statistical uncertainty, $(|\bias_i|+\Delta\bias_i)$/3, as 
our estimate of the systematic uncertainty in bias,
and here also add one third of its uncertainty
due to MC statistics.
Uncertainties in charged particle
momentum and calorimeter energy scale
are evaluated by shifting
those scales by the appropriate amount
and repeating the analysis.
We quote a relative momentum scale
accuracy of 0.01\%~\cite{cleometa}
and a calorimeter energy scale 
uncertainty of 0.6\%~\cite{cleometa}
and use these values
for our 1$\sigma$ systematic variations.
Any deviation from ideal in momentum
or energy scale is substantially
damped by the four-momentum 
constraints, as is
evident from Table~\ref{tab:tablesys}:
the momentum or energy scale
1$\sigma$ uncertainties
induce, at most, $\approx$3 parts in 10$^5$
shift in $\etap$-mass scale.

  To investigate the effect of less-well-measured
events, we have repeated the analysis after tightening
the kinematic fitting restrictions on each of
the four kinematic fits per event from
$\chi^2/$d.o.f.$<$8 to $<$3,
losing about half of the original events. 
The overall statistically-weighted $\etap$ mass
changes by 0$\pm$47~keV, in which the
uncertainty is statistical, demonstrating
stability of the measured mass with respect to
the kinematic fit quality.

  After combining the $(\aDMi$$-$$\bias_i)$ values in 
Table~\ref{tab:tableres} using the quoted weights and
adding the $\metapz$ offset from above, our result is
$\metap=957.793\pm0.054\pm 0.036~{\rm MeV}$,
where the first error is statistical and the second is systematic. 
This result betters the precision of the world average
by nearly a factor of four and has a central value consistent with it. 
The next most precise single measurement 
has a factor of five larger uncertainty~\cite{duane}
and is more than thirty years old.
Three $\etap$ decay modes contribute to this 
result and are consistent with one another.
This measurement
brings the mass uncertainty to 
a level more comparable to that of $K^0$ or $\eta$.
Including this result along with other
recent mass measurements in the
mixing angle prediction of Eq.~(1) gives 
$\phi_P$=(41.461$\pm$0.008)$^\circ$.

We gratefully acknowledge the effort of the CESR staff
in providing us with excellent luminosity and running conditions.
This work was supported by
the A.P.~Sloan Foundation,
the National Science Foundation,
the U.S. Department of Energy,
the Natural Sciences and Engineering Research Council of Canada, and
the U.K. Science and Technology Facilities Council.

\clearpage

\begin{table}[t]
\setlength{\tabcolsep}{0.45pc}
\catcode`?=\active \def?{\kern\digitwidth}
\caption{For each 
decay mode,
the number of events $N$, the Gaussian width on the mass
distribution of those data events,
$\sigma$, in MeV, the values of $\aDM$, $\bias$ (from MC),
and the difference $\aDM$$-$$\bias$
(see text), in keV.
Uncertainties shown are statistical.
}
\label{tab:tableres}
\begin{center}
\begin{tabular}{lrrrrrr}
\hline
\hline
Mode    & $N$\ \ & $\sigma$\ \ \ & $\aDM$\ \ \ \ & $\bias$\ \ \ \ & $\aDM$$-$$\bias$\ \  \\
\hline
$A$ & 2697 & 3.46 & $-$71$\pm$~\,85 & $-$61$\pm$20  & $-$10$\pm$87 \\
$B$ & 1017 & 1.93 &$-$113$\pm$~\,69 & $-$141$\pm$13 &  28$\pm$~\,70 \\
$C$ &  203 & 2.51 &$-$9$\pm$205     & $-$54$\pm$35  &  45$\pm$208   \\
$D$ &  230 & 1.95 & $-$3$\pm$147    & $-$41$\pm$19  &  38$\pm$148   \\
\hline
\hline
\end{tabular}
\end{center}
\end{table} 

\begin{table}[t]
\setlength{\tabcolsep}{0.45pc}
\catcode`?=\active \def?{\kern\digitwidth}
\caption{ For each $\etap$ channel, 
systematic and statistical uncertainties
in $\metap$ (in keV) from the listed sources (see text);
where applicable the degree of variation of the source
level is given.
The sources marked with a dagger ($^\dag$)
are assumed to be fully correlated across all modes;
others are assumed to be uncorrelated. 
The final column combines the uncertainties
across all modes with the weights given in the text. 
}
\label{tab:tablesys}
\begin{center}
\begin{tabular}{lrrrrr}
\hline
\hline
Source & Variation  & $A$ & $B$ & $C$ & All\\
\hline
Fit mass window        &                   & 11 &  9 & 31 &  7 \\
$M_\psip$$^\dag$       & 34~keV            &  9 &  2 &  3 &  5 \\
$M_\jpsi$$^\dag$       & 11~keV            &  3 &  2 &  2 &  2 \\
Bias                   & $(|\bias_i| +\Delta\bias_i)/3$       & 27 & 51 & 30 & 27 \\
$p_{\pi^\pm}$ scale    & 0.01\%            & 28 & 17 & 25 & 15 \\
$E_\gamma$ scale       & 0.6\%             & 13 & 22 & 28 & 12 \\
$\meta$                & 24~keV            &    & 23 &    & 11 \\
Systematic Sum         &                   & 44 & 63 & 57 & 36 \\
Statistical            &                   & 87 & 70 & 208& 54 \\
\hline
\hline
\end{tabular}
\end{center}
\end{table} 

\end{document}